\documentclass[draft]{agujournal2019}

\usepackage{url} 
\pdfoutput=1
\usepackage{amsmath}
\usepackage{soul}
\graphicspath{{./figs/}}
\makeatletter
\def\input@path{{./figs_new/}}
\makeatother
\newcommand{\ig}[2]{\includegraphics[width = #1]{#2}}

\draftfalse

\journalname{JGR: Space Physics}

\begin{document}

\title{The May 2024 Storm: dayside magnetopause and cusps in simulated soft X-Rays}


%
%




\authors{J. Ng \affil{1,2}, L.-J. Chen \affil{2}, B. Burkholder\affil{2}, D. Sibeck \affil{2}, F. S. Porter \affil{2}, K. H. Pham \affil{3}, V. G. Merkin \affil{3}, H. Connor \affil{2}, J. W. Bonnell\affil{4}, S. Petrinec\affil{5}, Y. Zou\affil{3}, B. Alterman \affil{2}, G. Cucho-Padin\affil{2}}
\affiliation{1}{Department of Astronomy, University of Maryland, College Park, MD}
\affiliation{2}{NASA Goddard Space Flight Center, Greenbelt, MD}
\affiliation{3}{Applied Physics Laboratory, Johns Hopkins University, Laurel, MD}
\affiliation{4}{Space Sciences Laboratory, University of California, Berkeley, CA}
\affiliation{5}{Lockheed Martin Advanced Technology Center, Palo Alto, CA, USA}

\correspondingauthor{Jonathan Ng}{jonng@umd.edu}







\begin{keypoints}
\item Reconstructed X-Ray imaging show magnetopause location variations consistent with observations and simulations
\item The cusps in X-Ray images expand poleward from nearly  parallel to each other as the IMF $B_z$ reverses from southward to northward.
\item Intense northern over southern cusp asymmetry is distinguished in X-Ray and understood as due to sunward dipole tilt and the solar wind flow into the cusp.

\end{keypoints}

\begin{abstract}
The coronal mass ejection (CME) arriving at Earth on May 10, 2024 caused the most intense geomagnetic storm in the last two decades, and resulted in highly unusual magnetopause and cusp dynamics. We simulate soft X-Ray emission due to solar wind charge exchange with exospheric neutrals to image the global dayside dynamics, focusing on the impact of a dense CME current sheet during the storm main phase. The magnetopause moves inward to $\sim$ 4 R$_E$, and at the same time, the two cusps manifest as nearly parallel emission ridges in X-Ray. As the interplanetary magnetic field reverses, the cusp ridges move to higher latitudes for $\sim 10$ minutes after the reversal. The X-Ray emission can be detected by imagers to be flown on future missions  to provide a global picture of the magnetopause and cusps with quantitative determination of their locations.
 
\end{abstract}

\section*{Plain Language Summary}
X-rays are produced when heavy ions in the solar wind and neutral hydrogen in the vicinity of the Earth interact. When captured by imagers, these X-rays are able to provide a view of the global magnetospheric structure. On May 10, 2024, a coronal mass ejection (CME) arriving at Earth caused the most intense geomagnetic storm in the last two decades, pushing the boundary between Earth's dipole and the interplanetary magnetic field inwards to $\sim 4 R_E$. We use simulations to reconstruct how this would have appeared using X-ray diagnostics, which will be used on future missions such as SMILE.

\section{Introduction}
The May 2024 storm, also known as the Mother’s Day storm or the Gannon storm, is the most intense storm in the last two decades. During the storm, the negative peak of Sym-H reached $-518$ nT \cite{tulasi:2024}, compared to $-585$ nT of the Hydro-Quebec storm that caused a major power outage across Quebec in 1989 \cite{boteler1989}. The interplanetary shock arrived at $\sim$17 UT on May 10, 2024, marking the initial phase of the storm. The strongest current sheet in the CME solar wind is embedded in a high density region (peaking above 100 cm$^{-3}$) corresponding to the reversal of the interplanetary magnetic field (IMF) with dominant $B_y$ reversing from approximately $-60$ to 70 nT (Fig. \ref{fig:omniweb}) during the main phase of the storm. The impact of the IMF reversal and density pulse causes extreme variations (4300 nT) in geomagnetic fields \cite{ohtani:2025}. The IMF reversal at ~22:30 UT on May 10, Fig. \ref{fig:omniweb} spans more than 100 R$_E$ transverse to the Sun-Earth line, since both Wind and ACE observe similar IMF profiles at L1. This study focuses on the soft X-Ray perspective of the dayside magnetopause and cusp dynamics around the impact time of this large-scale dense CME current sheet. Soft X-Ray reconstruction for other intervals of the May 2024 storm has been performed to address the magnetopause position variations under changing IMF \cite{gong:2025}. 

Using soft X-rays to image regions of the global magnetosphere is the goal of several missions, including the recent Lunar Environment heliospheric X-ray Imager (LEXI) \cite{walsh:2024}, the upcoming Solar wind-Magnetosphere-Ionosphere Link Explorer (SMILE) \cite{wang:2025}, and the mission concept Solar-Terrestrial Observer for the Response of the Magnetosphere (STORM) \cite{sibeck:2023}. 
The reconstruction of the expected X-ray images from simulation data has been demonstrated in various works \cite{kuntz:2015,sun:2015,walsh:2016,sun:2019,connor:2021,matsumoto:2022,ng:2023xray,grandin:2024}, and has shown the plausibility of observing magnetopause motion, flux-transfer events,  magnetosheath mirror-mode structures, and the bombardment of foreshock turbulence.

Following the modeling process described in detail in \cite{ng:2023xray}, we reconstruct X-ray emission around the impact time of the dense CME current sheet using a soft X-Ray model and output from a global geospace
simulation driven by spacecraft measurements. Our simulated soft X-Ray imaging provides the first global view of the dayside magnetopause and cusps driven by a dense CME current sheet. In Section~\ref{sec:model} we discuss the MHD
simulations and the X-ray emission model used to calculate the X-ray intensities. In Section~\ref{sec:results} we discuss the images produced and the observable physical phenomena, followed by a discussion of 
future possibilities and a summary in Section~\ref{sec:summary}.

\section{Soft X-ray model and global geospace simulation}
\label{sec:model}

In this work we use a previously established soft X-Ray production model \cite{cravens:2001, kuntz:2015, connor:2021} and data from global geospace simulations of the CME event to reconstruct soft X-Ray emission intensity driven by the interaction between solar wind heavy ions and the H in the geocorona.

\subsection{Soft X-ray model}

The X-ray intensity calculation is \cite{kuntz:2015,connor:2021, cravens:2001,ng:2023xray}:
\begin{equation}
  R_{xray} = \int \frac{\alpha}{4\pi} N_p N_N v_{\text{eff}}\, ds \left[\text{eVcm}^{-2}\text{s}^{-1}\text{sr}^{-1}\right],
  \label{eq:xray}
\end{equation}
where $v_{\text{eff}} = \sqrt{v_b^2 + v_{t}^2}$ is the effective velocity. $v_b$ is the bulk ion velocity while $v_{t}$ is the thermal velocity and $N_p$ is the ion density. Due to the nature of the simulation, these are MHD plasma quantities. The integral is line-of-sight and is taken along different angles in the field-of-view of the imager.

The parameter $\alpha$ is a scale factor that determines the soft X-ray production and includes the cross-sections, relative compositions, and transitions for heavy ion states that can give rise to soft X-rays, with $\alpha = 6\times 10^{-16}$ eVcm$^2$ for X-rays in the 100 eV to 1 keV range \cite{connor:2021,cravens:2001}. $N_N$ is the neutral density and we use a profile 
\begin{equation}
N_N = 25\left(\frac{10 R_E}{R}\right)^3\left[\text{cm}^{-3}\right],
\end{equation}
where $R$ is the distance from the centre of the Earth and $R_E$ is the Earth radius. 
The value of 25 cm$^{-3}$ is adopted by the SMILE Modeling Working Group for simulating  X-ray images and developing image analysis techniques \cite{connor:2025}.  Recent  studies on geocoronal and soft X-ray observations \cite{connor:2019, jung:2025, jung:2024, zoennchen:2021}showed that the subsolar hydrogen density at 10 $R_E$ is comparable to or higher than 25 cm$^{-3}$. Thus, 25 cm$^{-3}$ is a good conservative value.


\subsection{Global geospace simulation driven by observations}
The global geospace simulations are performed using the Multiscale Atmosphere-Geospace Environment (MAGE) model  \cite{lin:2021,pham:2022}. The MAGE configuration used in this study includes the inner magnetosphere model RCM \cite{toffoletto:2003}, ionospheric potential solver REMIX \cite{merkin:2010}, the thermosphere-ionosphere coupling model TIE-GCM \cite{richmond:1992}, and the global MHD model GAMERA \cite{zhang:2019,sorathia:2020}. The configuration of the model is similar to that used by \citeA{sorathia:2023} with the inner boundary at 1.5 $R_E$. The GAMERA warped spherical grid is aligned with the Solar Magnetic (SM) $X$-direction and  uses $96\times 96\times 128$ cells in the radial, azimuthal and polar directions respectively for simulating this event. The results presented in this work have been interpolated to a uniform Cartesian grid with $0.2 R_E$ resolution.  

\begin{figure}
  \centering
  \ig{3.775in}{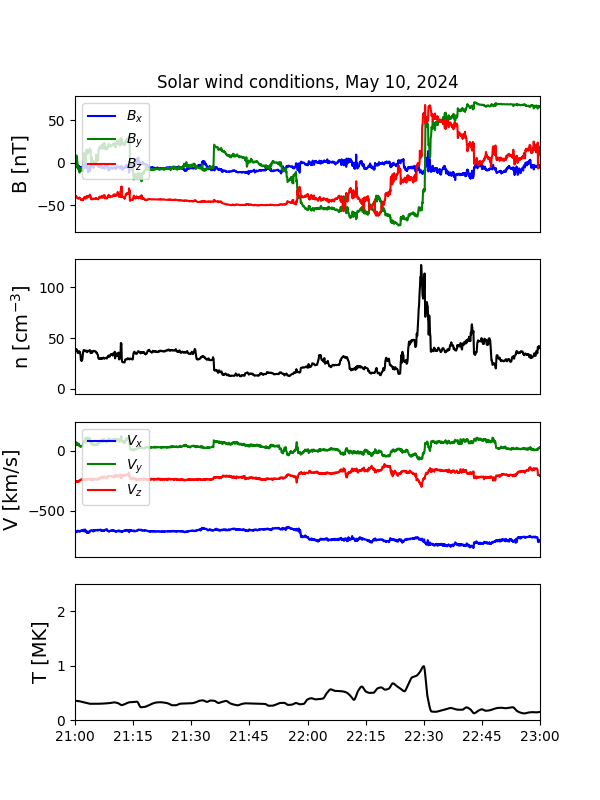}
  \caption{Upstream  magnetic field (B), density (n), velocity(V) and temperature (T) used as the simulation input. The data were measured by MMS (B, n, V) and Wind (T), and all data have been time-shifted to the bow shock nose. The reversal of the magnetic field $B_y$ and $B_z$ is associated with a density pulse reaching 120 $cm^{-3}$ and a temperature drop immediately after the density pulse.}
  \label{fig:omniweb}
\end{figure}

The MAGE simulation presented here is the same as previously used by \citeA{ohtani:2025}. For the solar wind and IMF conditions, we use a combined upstream MMS (B, n, V) and Wind (T) data set as input (Fig.~\ref{fig:omniweb}).
We focus on the interval that encompasses the density pulse associated with the reversal of the IMF $B_y$ and $B_z$ at approximately 22:30 UT.

\section{Results}
\label{sec:results}

\begin{figure}
  \centering
  \ig{5.375in}{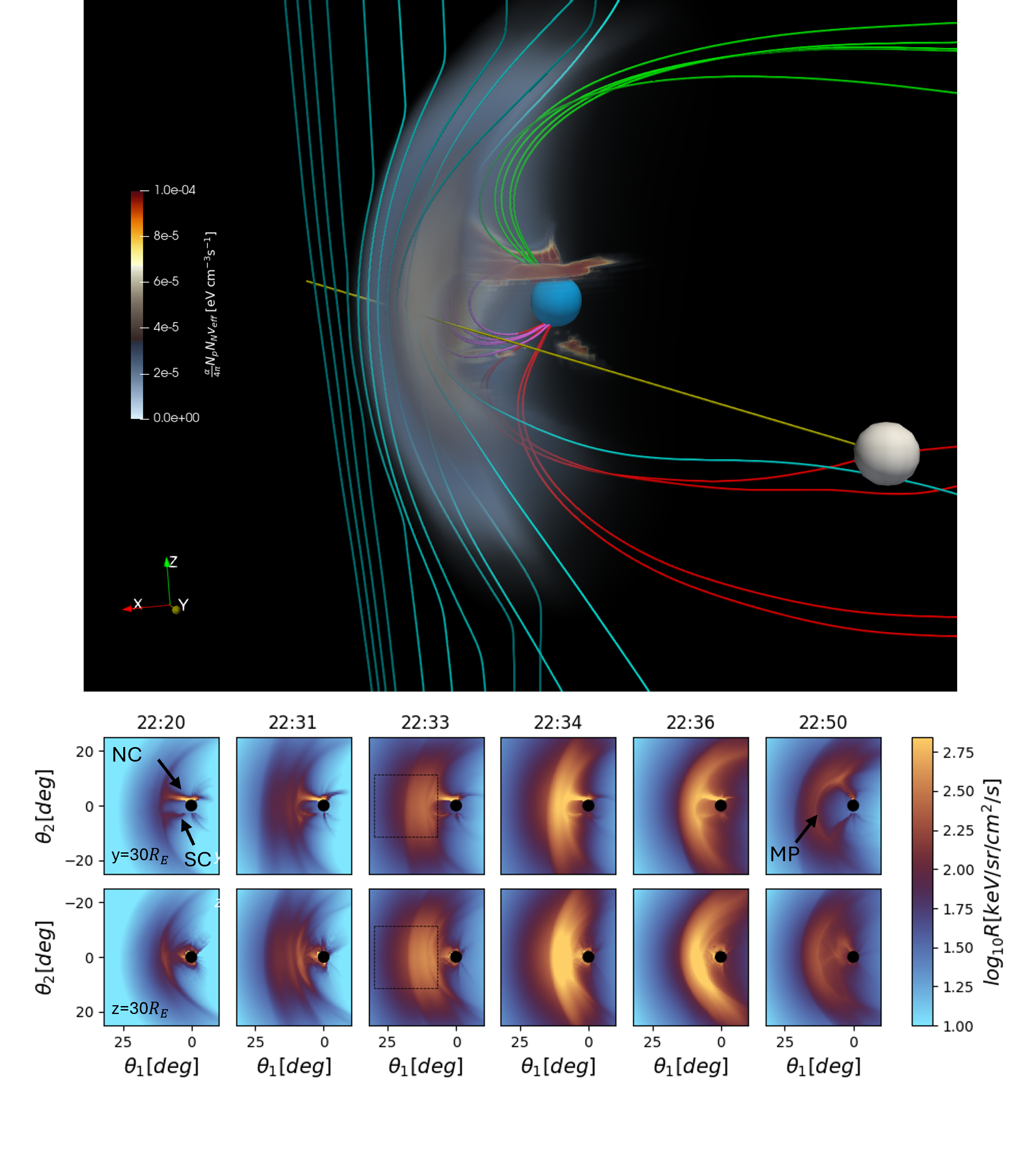}        
  \caption{(Top) Volume rendering of the integrand of Equation~\eqref{eq:xray} at 22:36 UT. The white sphere and yellow line show a line-of-sight from a spacecraft positioned at $y=30 R_E$ ($x=z=0$). IMF field lines are drawn in cyan. Red and green field lines connect to the southern and northern cusps respectively. Magenta field lines are closed. (Bottom) Soft X-ray intensity after line of sight integration observed by an imager at [0, 30, 0] $R_E$ (top row) and at [0, 0, 30] $R_E$ (bottom row) . The  soft X-ray images are constructed with an angular resolution $0.25^\circ \times 0.25^\circ$. The STORM mission is designed to have an angular resolution $0.17^\circ$ to $0.25^\circ$ and a field of view marked by the boxes. Annotations show Northern cusp (NC), Southern cusp (SC) and magnetopause (MP). All vectors are in the SM coordinates.}
  \label{fig:3d}
\end{figure}

 In this Section, we present the dayside magnetosphere in soft X-Ray with the X-ray intensity in an example 3D volume rendering and using the results of line of sight integration. 
The X-Ray intensity at each pixel at 22:36 UT is visualized together with a selection of magnetic field lines (Figure \ref{fig:3d}, top). IMF field lines are drawn in cyan. Red and green field lines connect to the southern and northern cusps respectively while the magenta field lines are closed. The X-ray signal is strongest in the cusps and in the magnetosheath. The sphere placed at (0, $y = 30 R_E$, 0) illustrates a possible position of the imaging spacecraft (taken as the STORM mission) and the yellow line illustrates the line-of-sight along which an integral is taken.
The 3D volume rendering represents the integrand of equation~\eqref{eq:xray}, calculated using the output of the global simulation. The signal in the closed field-line region is set to zero assuming that the high-charge state ions cannot enter the dayside closed-field-line region within the time scale of our study \cite{sibeck:2018}. A line-of-sight integral is then performed from the imager location. We use the angular resolution of the X-Ray imager in the STORM mission concept \cite{sibeck:2023,murphy:2024}, designed to study the global solar wind-magnetosphere interaction.

The reconstructed X-ray images of the dayside magnetopause and cusps reveal the 3D magnetosphere reconfiguration around the time of the dense CME current sheet arrival.
We follow the evolution using spacecraft placed at (0, $y = 30 R_E$, 0) and (0, 0, $z = 30 R_E$) as shown in Fig.~\ref{fig:3d}. These are planned locations of the STORM mission, which is intended to have a $30 R_E$ orbit perpendicular to the ecliptic plane. The angular resolution of the presented data is $0.25^\circ \times 0.25^\circ$, the intended resolution of the STORM X-Ray imager.  Here $\theta_1$ and $\theta_2$  define the viewing angles, with $\theta_1$ corresponding to the looking direction towards different $X$, and $\theta_2$ the north-south (the ``$y=30 R_E$" row) or dawn-dusk (the ``$z=30 R_E$" row) directions.

The arrival of the density pulse can be seen from images at 22:31 to 22:34 UT, with multiple local maxima in the intensity at positive $\theta_1$ corresponding to the density peaks modulated by the $1/R^3$ structure of the neutrals. The innermost intensity peak corresponds to the magnetopause position, where the line of sight viewed from $y=30R_E$ is tangent to the subsolar magnetopause surface. There is an approximate fourfold increase of the peak intensity from 22:31 to 22:34 UT along $\theta_2 = 0$ in the magnetosheath. The magnetopause moves toward Earth from 22:20 to 22:34 UT with the closest distance at $\sim 4 R_E$ (Figure 3) as the dense current sheet (IMF reversal and density pulse) arrives, and the magnetopause expands away from Earth from 22:34 to 22:50 UT as the dense current sheet passes the magnetopause.

The two cusps manifest themselves as two emission ridges in X-Ray with the northern cusp substantially brighter than the southern cusp (Figure \ref{fig:3d}, top row of thumb-nail panels). This asymmetry is due to the sunward dipole tilt as well as the southward component of the solar wind flow leading to higher solar wind flux into the northern cusp. The two ridges are nearly parallel to each other and the sun-earth line until $\sim$ 22:33 UT, approximately three minutes after the IMF reversal from southward to northward. Thereafter, the cusp X-Ray emission ridges move
poleward, as the magnetopause retreats farther away from Earth.
The nearly parallel cusp ridges and the pushed-in magnetopause location
indicate that the dayside magnetic flux (closed field lines) is highly eroded. The northward turning of the IMF starts magnetic reconnection poleward of the cusps, and consequently the two cusp ridges move poleward and away from each other. 
X-Ray imaging provides a global view of the magnetic flux erosion and re-generation process, and the possibility to quantify the variation of the flux content as well as the time scale. 
For example, the addition of flux to the dayside can be seen through both the displacement of the magnetopause in the positive $\theta_1$ direction as well as the cusp moving poleward. The area corresponding to the closed-field-line-region (bounded by the two cusps and the magnetopause) of the magnetosphere in the reconstructed images is approximately four times larger at 22:50 UT compared to that at 22:34 UT. 



\begin{figure}
  \centering
  \ig{5.75in}{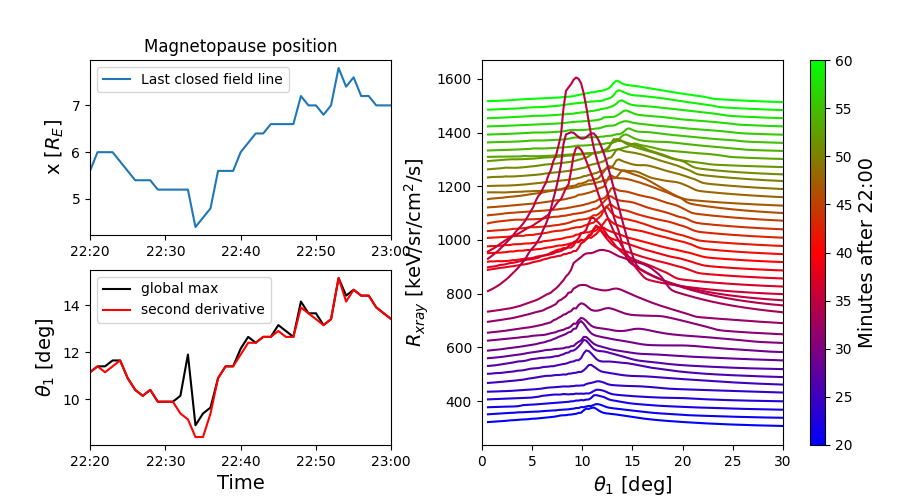}
  \caption{[Left] Magnetopause position determined by magnetic topology (top panel) and extrema in the X-ray intensity and its second derivative (bottom). [Right] X-ray intensity as seen from $y = 30 R_E$ along $\theta_2 = 0$ showing the motion of the magnetopause. }
  \label{fig:mp_position}
\end{figure}

As mentioned earlier, the magnetopause position can be identified by finding the peak intensity along $\theta_1$, or a peak in its second derivative \cite{connor:2021,collier:2018}. The motion of the magnetopause is illustrated in Figure~\ref{fig:mp_position}. The top-left panel shows the location of the last closed field line along $x$ at $y=0, z=0$ in the simulation, and there is abrupt inward displacement of the magnetopause to within 4-5 $R_E$ radial distance at around 22:34 UT, followed by relaxation. This unusually short distance of the magnetopause standing off the CME wind is consistent with observations by THEMIS (constrained to be $< 6 R_E$ \cite{fu:2025}) and Arase ($\sim 5-6 R_E$ \cite{miyoshi:2025}) spacecraft.  The bottom left panel shows the magnetopause position inferred from the X-ray intensity using either the maximum intensity or the minimum of $d^2R_{xray}/d\theta_1^2$. Both methods produce general agreement with the result from magnetic topology, even though there is a  discrepancy around the arrival time of the density pulse, with the maximum intensity method showing a more sunward magnetopause position. The cause is illustrated in the right panel, which shows the variation of intensity along $\theta_1$ at $\theta_2 = 0$, in which the colours represent time. The arrival of the CME current sheet can be seen in the increase of the X-ray intensity sunward of the magnetopause (larger $\theta_1$), with the maximum intensity observed due to the dense current sheet rather than the magnetopause boundary at certain times. More complex methods exist that extract the magnetopause location in GSE coordinates from soft X-ray emission peak angles and determine the shape of magnetopause (e.g.~\citeA{collier:2018, kim:2023,cuchopadin:2024, gong:2025}).


What would the STORM soft X-Ray imager observe under extreme solar wind conditions such as the May 2024 storm? In Figure~\ref{fig:3d}
we show the expected results using the field of view of the proposed STORM imager , which is $23^\circ \times 23^\circ$ field of view pointed $18.5^\circ$ sunward of nadir, and is illustrated by the dashed box in Figure~\ref{fig:3d}. STORM will be able to track the magnetopause position over time, as the density pulse pushes the subsolar magnetopause closer and closer to Earth up to $\sim 4 R_E$ radial distance. 
At 22:34 UT, when the magnetopause position reaches $X \sim 4-5 R_E$, the subsolar magnetopause is still visible, although it is close to the Earthward edge of the planned field of view of the STORM imager.  

The question of whether this X-ray signal can be measured is important. The results thus far are instantaneous, whereas an actual instrument would have finite time integration and be affected by backgrounds. During the arrival of the CME current sheet, the peak intensity in the magnetosheath is over 800 keV cm$^{-2}$s$^{-1}$sr$^{-1}$, compared to the diffuse soft X-ray background of 10-30 keV cm$^{-2}$s$^{-1}$sr$^{-1}$ \cite{sibeck:2018}. Based on previous work on reconstructions \cite{walsh:2016},  this unusually high signal from the magnetosheath, because of the high density, would enable high cadence (sub 1 minute) images  to be captured.



\section{Summary and Discussion}
\label{sec:summary}

Our simulated soft X-Ray images reveal the global dynamics of the dayside magnetopause and cusps during the May 2024 storm, particularly in response to prolonged strong southward IMF and an unusually dense CME current sheet.
Using the MAGE global simulations of the May 10 2024 CME event, we  reconstruct the X-Ray images expected to be measured by spacecraft located at (0, $y = 30 R_E$, 0) and (0, 0, $z = 30 R_E$). The combined X-Ray and global simulation results show that (1)
the angular position of the magnetopause moves earthward during the arrival of the CME sheet, consistent with spacecraft observations and field-line-tracing simulation results, (2) the cusps in X-Ray emissions move poleward from nearly  parallel to each other with a time delay a few minutes after the IMF $B_z$ turns from southward to northward, and the migration to higher and higher latitudes occurs over $\sim 10$ minutes, and (3) the northern cusp is substantially brighter in X-Ray than the southern cusp. 


The global X-Ray view of the magnetopause and cusps complements those from the ionospheric measurements.
Based on ionospheric radar measurements from the SuperDARN network \cite{greenwald:1995}, the open-close magnetic field boundary (the low-latitude boundary of the cusp) at $\sim$ 22:20 UT  is at MLAT $\sim 56^\circ$, and at 22:50 UT, it is further poleward between $62^\circ$ and $66^\circ$.
 Using a composite model of the cusp location and extent at ionospheric altitudes (constructed from empirical relations of dependences on solar wind conditions and the dipole tilt angle in the literature) \cite{petrinec:2023} for a southward IMF $B_z=-50$ nT (such as the $B_z$ at 22:20 UT, Figure 1) and a dipole tilt of 18.3$^\circ$, the equatorward edge of the northern cusp is obtained to be 58-59$^\circ$, consistent with the radar observation. When the IMF is northward with $B_z\sim 50$ nT (just after the large IMF $B_y$ rotation, Figure 1) the equatorward edge of the northern cusp is at $\sim 85^\circ$ based on the model.
 The X-Ray picture shows the large-scale cusp structure in a radial distance range $\sim 1.5 R_E$ to $\sim 5 R_E$, and its evolution from approximately perpendicular to forming an acute angle with respect to the dipole axis (along positive z in the SM coordinate used in this paper), information not available from the ionospheric measurements or the empirical cusp model.

Soft X-ray imaging can provide simultaneous observations of the impact of solar-wind interactions on different regions of the magnetosphere. As such, it complements in situ measurements of the magnetospheric plasma by providing the global context, such as the locations and structures of the magnetopause, cusps, the bow shock, and magnetosheath. The global X-ray images from future missions will provide large-scale information that is challenging, if not impossible, for even multi-point local measurements to deliver.

\acknowledgments
The work of JN, LJC, and BB was supported by the NASA MMS mission, and the work of KHP and VGM by the NASA DRIVE Science Center for Geospace Storms (CGS) under NASA award 80NSSC22M0163. The MAGE model is developed by CGS. We acknowledge high-performance computing support from NASA High-End-Computing and the Derecho system (doi:10.5065/qx9a-pg09) provided by the NSF National Center for Atmospheric Research (NCAR), sponsored by the National Science Foundation.

\section*{Open Research}
Data used in this paper are openly available at \cite{ng:2025_data}.

\bibliography{reconnectionbib.bib}

\end{document}